\documentclass[aps,pre,superscriptaddress,twocolumn,nofootinbib]{revtex4-1}
\usepackage[dvipsnames]{xcolor}
\usepackage{amsmath,amssymb}
\usepackage{graphicx}
\usepackage{grffile} 
\usepackage{footmisc}
\usepackage{hyperref}
\usepackage{bbold,soul}
\usepackage{lipsum}

\makeatletter
\newcommand\footnoteref[1]{\protected@xdef\@thefnmark{\ref{#1}}\@footnotemark}
\makeatother

\newcommand{\pd}{P_\downarrow}
\newcommand{\pu}{P_\uparrow}
\newcommand{\pz}{P_0}
\newcommand{\bc}{\beta_\mathrm{c}}
\newcommand{\tw}{t_\mathrm{w}}
\newcommand{\rth}{r_\mathrm{th}}
\newcommand{\Eth}{E_\mathrm{th}}

\newcommand{\vecx}{\mathbf{x}}

\newcommand{\vecn}{\mathbf{n}}

\newcommand{\code}[1]{\texttt{#1}}

\begin{document}
\title{Effective Trap-like Activated Dynamics in a Continuous Landscape}
\author{Matthew R. Carbone}
\email[]{mrc2215@columbia.edu}
\affiliation{Department of Chemistry, Columbia University, New York, NY 10027, USA}
\author{Valerio Astuti}
\affiliation{Dipartimento di Fisica, Sapienza University of Rome, P. le A. Moro 5, Rome 00185, Italy}
\author{Marco Baity-Jesi}
\affiliation{Department of Chemistry, Columbia University, New York, NY 10027, USA}
\affiliation{Eawag, \"Uberlandstrasse 133, CH-8600 D\"ubendorf, Switzerland}

\date{\today}

\begin{abstract}
We use a simple model to extend
network models for activated dynamics to a continuous landscape with a well-defined notion of distance and a direct connection to many-body systems.
The model consists of a tracer in a high-dimensional funnel landscape with no disorder. We find a non-equilibrium low-temperature phase with aging dynamics that is effectively equivalent to that of models with built-in disorder, such as Trap Model, Step Model and REM.
Finally, we compare entropy- with energy-driven activation, and we remark that the former is more robust to the choice of the dynamics, since it does not depend on whether one uses local or global updates.
\end{abstract}

\maketitle

\section{Introduction}
Glasses display a extraordinarily slow dynamics as temperature is decreased~\cite{debenedetti:97}. 
The mean-field picture of this slowing down is elegantly explained as a topological transition in the energy landscape: while at high temperature $T$ the typical configurations are close to the saddle points of the energy landscape, under the \emph{dynamical temperature} $T_\mathrm{d}$ the system is confined near the minima of the landscape. Since in mean-field models the energy barriers $\Delta E$ diverge in the thermodynamic limit, for $T<T_\mathrm{d}$ the system remains confined near the local energy minima and ergodicity is broken~\cite{cavagna:09}.

In non-mean-field systems, the barrier heights remain finite, and can be overcome in time scales $\tau$ that follow the Arrhenius law, $\tau\propto\exp(\frac{\Delta E}{k_\mathrm{B}T})$~\cite{kurchan:11}.\footnote{From here on we set the Boltzmann constant  $k_\mathrm{B}=1$ and note that unless written with a specified base, all logarithms are
the natural logarithm.}
This barrier-hopping dynamics, which correspond to collective rearrangements of particles, are also called \emph{thermally activated}. The ergodicity breaking induced by the topological transition in the mean field model is hence avoided in many systems of interest, such as 3D glass formers. Activated dynamics must thus be understood in order to characterize the slowing down of glasses.

Given the overwhelming difficulties in the theoretical description of low-dimensional glass formers, a first step towards the understanding of activated dynamics should be done in the mean field approximation. Barrier crossing is in fact also possible in the mean field approximation, provided that the system size $N$ is kept finite~\cite{crisanti:00,crisanti:00b}. Keeping $N$ finite makes calculations especially hard, so save for some exceptions~\cite{benarous:02,gayrard:16,gayrard:18}, most work on activated dynamics consists of numerical simulations~\cite{crisanti:04b,heuer:08,stariolo:19}. 

The most popular theoretical framework for the interpretation of activated dynamics is the Trap Model (TM)~\cite{dyre:87,bouchaud:92}, which consists of a simplified, solvable, version of the energy landscape of glasses, in which the only way to explore the phase space is a purely activated motion between minima in the landscape (called traps), with no notion of distance and with a fixed \emph{threshold} energy that needs to be reached in order to escape a trap. 

The TM yields a wide set non-trivial quantitative and qualitative predictions that have been used to rationalize numerical simulations of low-dimensional glass formers~\cite{denny:03,doliwa:03,heuer:08}, and has been recently shown to serve as an accurate representation of the dynamics of some simple models of glasses in which a threshold energy can be easily identified~\cite{gayrard:16,gayrard:18,baityjesi:18}.\footnote{{This connection between trap models had been anticipated decades earlier (see e.g. Refs.~\onlinecite{benarous:02} and \onlinecite{dyre:95}).}}
A TM description of the dynamics was also shown to be accurate in the Step Model (SM), a model with a single energy minimum, provided that one identifies traps in a dynamical way~\cite{cammarota:15}. 

Despite these successes, the TM suffers from limitations. On one side, it pictures a phase space motion that is completely unrelated to real-space degrees of freedom, and it is defined on a discrete space of configurations. 
The first of these two issues was successfully addressed by showing that some problems from number theory can be reformulated as physics problems on a lattice, which behave like the TM~\cite{junier:04,junier:04b}.
On the other side, the TM paradigm of activated dynamics is probably not suitable for the description of most systems with strong enough correlations~\cite{baityjesi:18c,stariolo:19,baityjesi:20}. 
One must therefore try to understand the limits of the applicability of the TM paradigm, and whether it is possible to create a connection between the TM and other systems such as sphere packings.

Much progress along these lines was made in a series of works culminating with the proof that the Random Energy Model (REM), a simple model with glassy behavior, exhibits trap-like dynamics~\cite{benarous:02,benarous:06,gayrard:16,gayrard:18,gayrard:18b}.
Another consists of studying the influence of phase space connectivity on the dynamics~\cite{margiotta:18,margiotta:19}. In our approach,
we show that the TM paradigm also applies to a very simple model of a \emph{continuous} $N$-dimensional landscape, where each dimension represents an independent coordinate in a fictitious space with well-defined metrics. This is done by noting that a TM-like activated behavior can arise 
due to entropic effects also in the absence of multiple local minima, as was shown for the SM~\cite{cammarota:15}.

Our work is thus organized as follows:
In Sec.~\ref{sec:prelim} we make simple preliminary observations on how dimensionality induces entropic effects, and
in Sec.~\ref{sec:model} we introduce the physical model. Furthermore, we study its out-of-equilibrium behavior in Sec.~\ref{sec:off-eq} and in Sec.~\ref{sec:trap} we show that TM-like dynamics arise. Finally, in Sec.~\ref{sec:conc} we summarize and discuss our results. We also provide a discussion of the details of our numerical simulations, and mathematical derivations,
in the appendices.

\section{Dimensionality and Entropic Effects}\label{sec:prelim}
We study the dynamics of a tracer in an $N$-dimensional space. This is meant to represent the phase space dynamics of a many-body system in a central potential.\footnote{Note that these are $N$ one-dimensional interacting particles, or a single particle in an $N$-dimensional space.}
The potential energy depends only on the distance $r$ of the tracer from the origin,
\begin{equation}
    \tilde E(r) = \log r,
\end{equation}
and we restrict the phase space to $r \in (0, 1]$.
Even though it seems clear that a steepest descent minimization would lead to the origin, any source of noise (temperature, step size, numerical, etc.) would make it almost surely unreachable.

\begin{figure}[!htb]
\includegraphics[width=1.0\columnwidth]{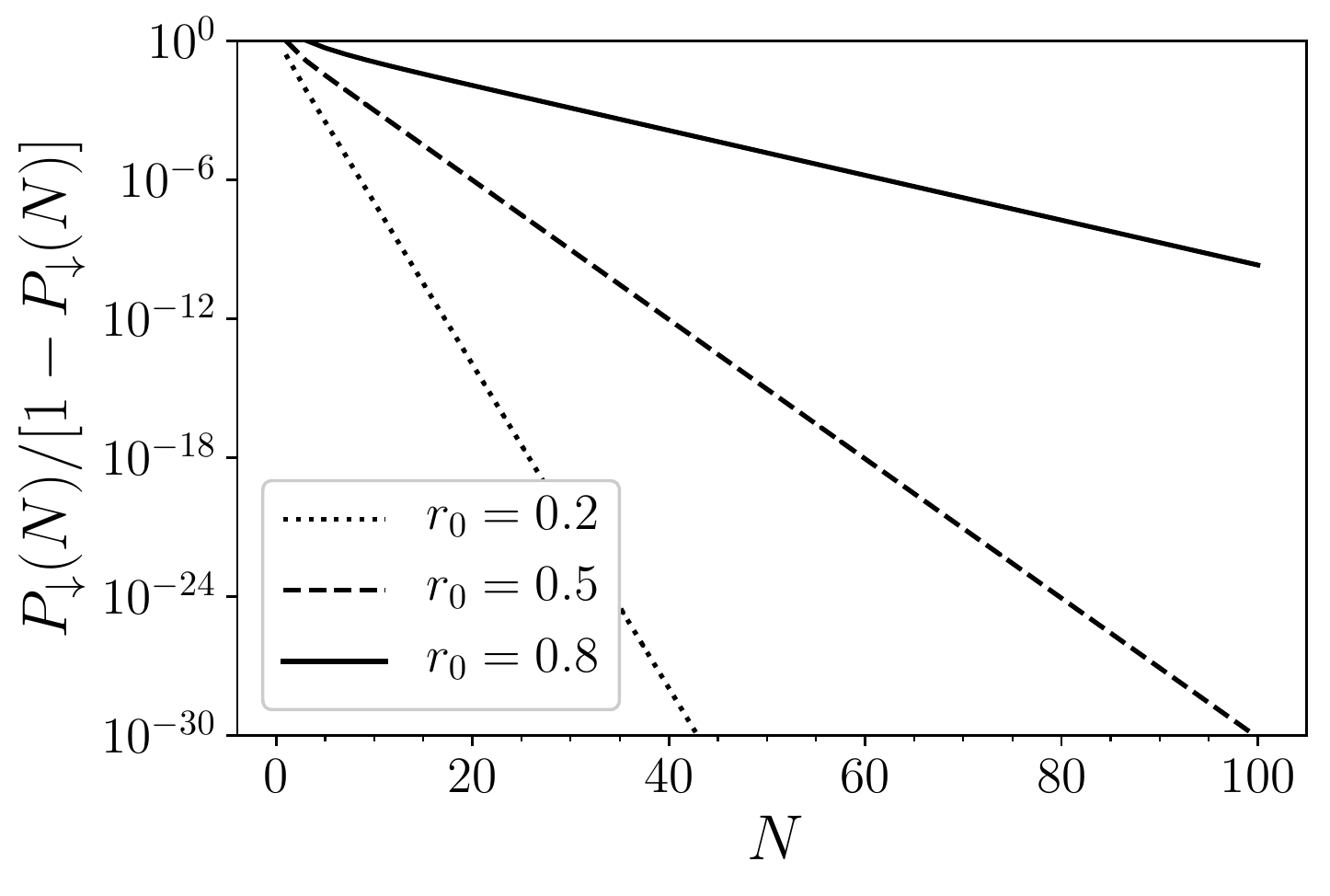}
\caption{\label{fig:pdown} Ratio of the probability of the tracer moving towards the well singularity, $\pd,$ and the
tracer moving either away from it or staying in the same location, $1-P_{\downarrow}$. Each curve represents a different starting radius, $r_0$.}
\end{figure}

When evolving through a generic equilibrium dynamics algorithm, if $N$ is large, the system would barely feel the
presence of the energy funnel, despite its negative divergence. As an example, let us take a Monte
Carlo direct sampling dynamics on the $N$-dimensional unit hypersphere. At every time step a
new configuration is proposed with a
uniform probability, and a transition towards it is accepted with probability
\begin{equation}\label{eq:pmc}
    p_\mathrm{MC} = \min(1,e^{-\beta\Delta E})\,,
\end{equation}
where $\beta = T^{-1}$ is the inverse temperature and the energy difference $\Delta E$ is negative
if the transition decreases $r$.
From any position $\vecx_0=(x_{0,1},\ldots,x_{0,N})$ in the landscape, the probability $\pd$ of moving towards lower energy is set by the relative volume of the sphere of radius $r_0=|\vecx_0|$,
\begin{equation}\label{eq:pdown}
\pd(r_0) = \frac{V_N(r_0)}{V_N(1)} = r_0^N\,,
\end{equation}
where $V_N(r)$ is the volume of the $N$-dimensional sphere of radius $r$. As also depicted in
Fig.~\ref{fig:pdown}, the probability of proposing a move that decreases the energy goes down
exponentially with the dimension of the system $N$, and this decrease is more severe the closer $\vecx_0$ is to the origin. In Fig.~\ref{fig:pdown} we can remark that, already for dimensions as small as $N=10$, $\pd$ is smaller than single floating point accuracy.

From Eq.~\eqref{eq:pdown} we can extract a characteristic time scale for decreasing the energy
\begin{equation}\label{eq:tau0}
    \tau_0\sim1/\pd=e^{-N\log r_0}\,.
\end{equation}
This kind of slowing down generated by the rarefaction of directions that decrease the energy is called \emph{entropic aging}~\cite{barrat:95,bertin:10}.

At non-zero temperature, for large-enough $N$ the dynamics will always be pushed outwards, because the probability of
accepting moves will stay finite (i.e. independent from $N$), while the probability of proposing moves that decrease the energy is dramatically
suppressed. In other words, unless the potential energy also scales with $N$, at any fixed $r_0$ there will always exist an $N$ over which the attraction that $\tilde{E}(r)$ exerts on the tracer becomes irrelevant.
One therefore needs to counterbalance by either taking temperatures of order $1/N$, or giving the potential the right scaling $N$, in order to account for the
energetic push towards the center of the $N$-dimensional sphere with the entropic effects induced by large dimensionality that push the system towards the boundary. This is standard practice in statistical mechanics problems.

\section{Physical Model}\label{sec:model}
If we recall that our model describes the phase space of an $N$-body system, we should not be surprised that plausible
dynamics require the potential to be rescaled with $N$, as the energy should be an extensive quantity.
Thus, in the rest of this work we will use the properly re-scaled potential
\begin{equation}
     E(r)=\frac{N}{\bc}\log r \,,
\end{equation}
where $\bc>0$ is a parameter that sets the right physical dimensions, and $r \in (0, 1]$.
We normalize the volume so that the portions of phase space map exactly onto their probability, in agreement with a microcanonical interpretation of the space of configurations. The normalized radial volume element is then $d \tilde{V}=\frac{dV}{V_N(1)}=\frac{\Omega^N N r^{N-1}d r}{V_N(1)} = Nr^{N-1}d r = \bc e^{\bc E}dE$, where $\Omega^N$ is the $N$-dimensional solid angle.
As a consequence, the density of states $g(E)\equiv|\frac{d\tilde{V}}{dE}|$ is equal to
\begin{equation}\label{eq:dos}
    g(E)= \bc e^{\bc E}\Theta(-E)\,,
\end{equation}{}
where $\Theta(x)$ is the Heaviside step function.

From $g(E)$ we can calculate the partition function of the Canonical Ensemble,
\begin{equation} \label{eq:partition_function}
    Z(\beta; \beta_\mathrm{c})=\int_{-\infty}^0 \bc e^{(\bc-\beta)E}dE = \frac{\bc}{\beta-\bc}\,,    
\end{equation}
which is well-defined only for $\beta<\bc$. Thus, the equilibrium phase, with average energy $\langle E(\beta)\rangle = \frac{1}{\bc-\beta}$ and radius $\langle r(\beta)\rangle = \exp\left(\frac{\bc}{N(\beta-\bc)}\right)$, only exists for $\beta\leq\bc$. For $\beta>\bc$, the system is out of equilibrium and the energy will eventually diverge to $-\infty$ as time goes to infinity.

\section{Off-equilibrium Dynamics}\label{sec:off-eq}
Although the equilibrium phase is trivial, the out-of-equilibrium phase of this model displays rich behavior. Since we are out of equilibrium, we need to define the kind of dynamics used: we analyze both non-local (global) and local dynamics, which are generally equivalent for equilibrium simulations. For global dynamics we use Direct Sampling Monte Carlo
(DSMC), and for local dynamics, Markov Chain Monte Carlo (MCMC).
Details on simulations and measurements are given in Appendix~\ref{app:sim}.

\subsection{Direct Sampling Monte Carlo}
With DSMC, at every time step a point in the hypersphere is proposed as a move for the algorithm. All points of the phase space are proposed with equal probability, and the moves are accepted with probability $p_\mathrm{MC}$ [Eq.~\eqref{eq:pmc}].

In the following, we show that according to the value of $\beta$ there are several regimes in the dynamics, which were already found in the SM. For $\beta>2\bc$ the energy decreases slowly and steadily, at a rate that follows Eq.~\eqref{eq:tau0}; we call this the entropic aging (EA) regime~\cite{barrat:95,bertin:10}. For intermediate $\beta$, even though the energy as a function of time is decreasing on average,\footnote{We mean an average over the trajectories, not an ensemble average, which is not well defined for $\beta>\bc$. We use an overbar, $\overline{(\ldots)}$, to denote average over trajectories.} the trajectory intermittently returns to high energies. Following previous literature on the SM, we call this regime \emph{thermally activated}~\cite{bertin:03,bertin:12,cammarota:15}.
In this regime, one can identify a finite \emph{threshold} energy (or, equivalently, radius) towards which the dynamics is spontaneously driven.\footnote{This concept of threshold energy is related to its definition in the TM, as the energy to reach in order to jump barriers and have renewal dynamics~\cite{bouchaud:92}. Operative methods for the detection of the threshold energy based on the behavior of $p$-spin-like models (e.g. those used in Refs.~\onlinecite{hartarsky:19,stariolo:19}) are not good methods in this context.
} Even though the dynamics intermittently returns to the threshold, $\overline{E}(t)$ is decreasing because the system spends short times at high energy, and increasingly longer times at lower energy.

Following Refs.~\cite{bertin:03,cammarota:15}, we define the threshold radius $\rth$ as the radius from which the probability $\pu$ of increasing the energy equals the probability of decreasing it,
\begin{equation}
    \pu(\rth;\beta)\equiv\pd(\rth)\,.
\end{equation}
With Monte Carlo dynamics, in general $\pu+\pd<1$, since there is also a non-zero probability $\pz$ that the tracer does not move due to the rejection of movement proposals. However, our calculations of $\rth$ are static, so $\pz$ does not influence them. Neglecting $\pz$ in a dynamic calculation is equivalent to saying that time does not advance when a move is rejected: this does not change the probability of increasing or decreasing the energy once the move gets accepted.

\begin{figure}[!htb]
\includegraphics[width=1.0\columnwidth]{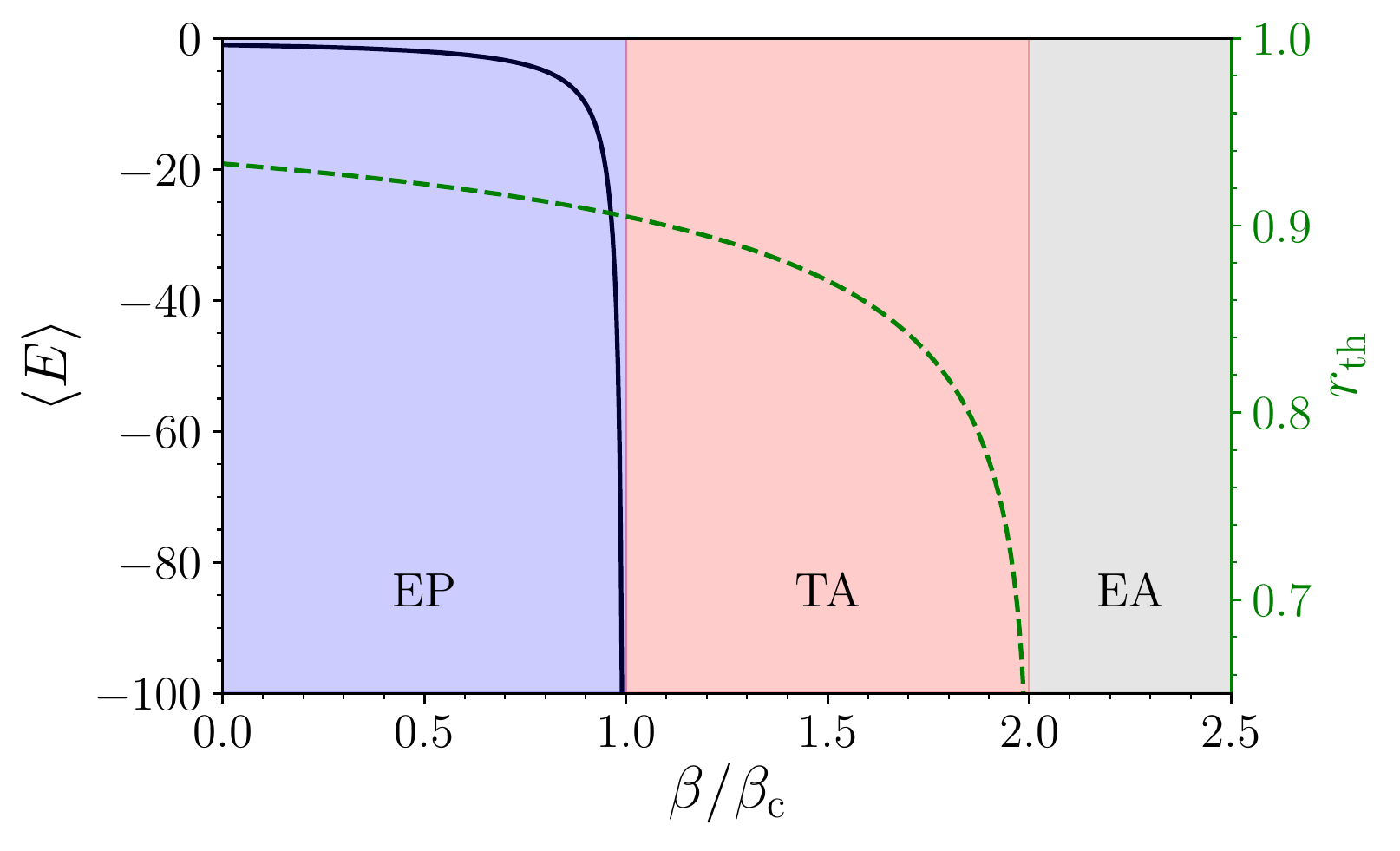}
\caption{\label{fig:DSMC_rth_expE}
Threshold radius (green) and equilibrium value of the energy (black)
plotted as a function of $\beta/\bc$ for $N=10$. The equilibrium phase (EP) regime, where the system enjoys both a well-defined threshold radius and equilibrium energy
is shown in blue ($0 < \beta < \bc$). The thermal activation (TA) regime
is shown in red. Finally, the entropic aging (EA) regime is
shown in grey, where the tracer is relentlessly attracted towards the center of the well ($\beta > 2 \bc$).}
\end{figure}

The probability of increasing the energy from a radius $r_0$ with DSMC is
\begin{align} \label{threshold radius}
    \pu(r_0;\beta) =\,& \frac{\Omega^N}{V_N(1)}\int_{r_0}^1 r^{N-1} e^{-N\frac{\beta}\bc(\log r-\log r_0)} dr =\\
        =\,& \frac{r_0^N-r_0^{N\beta/\bc}}{\frac\beta\bc-1}\,,
\end{align}
while $\pd(r_0)$ is given in Eq.~\eqref{eq:pdown}. Equating the two, one obtains no real solution for $\beta>2\bc$. For $\bc\leq\beta\leq2\bc$ there is a solution growing continuously from 0 at $\beta=2\bc$ to 1 at $\beta=\bc$,
\begin{equation}\label{eq:rth}
    \rth\left(\beta \right) = \left(\frac{2\bc-\beta}{\bc}\right)^\frac{\bc}{N \left(\beta-\bc \right)}\,.
\end{equation}
For $\beta<\bc$ we are in the equilibrium phase: $\rth$ is at distance $\sim1/N$ from the system boundary and from $\left\langle r\right\rangle$.
Summarizing, in our simple funnel model we have three regimes (Fig.~\ref{fig:DSMC_rth_expE}):
\begin{itemize}
    \item $\beta<\bc$: Equilibrium Phase (EP)
    \item $\bc\leq\beta<2\bc$: Thermal Activation (TA)
    \item $\beta\geq2\bc$: Entropic Aging (EA)
\end{itemize}

\begin{figure}[!htb]
\includegraphics[width=1.0\columnwidth]{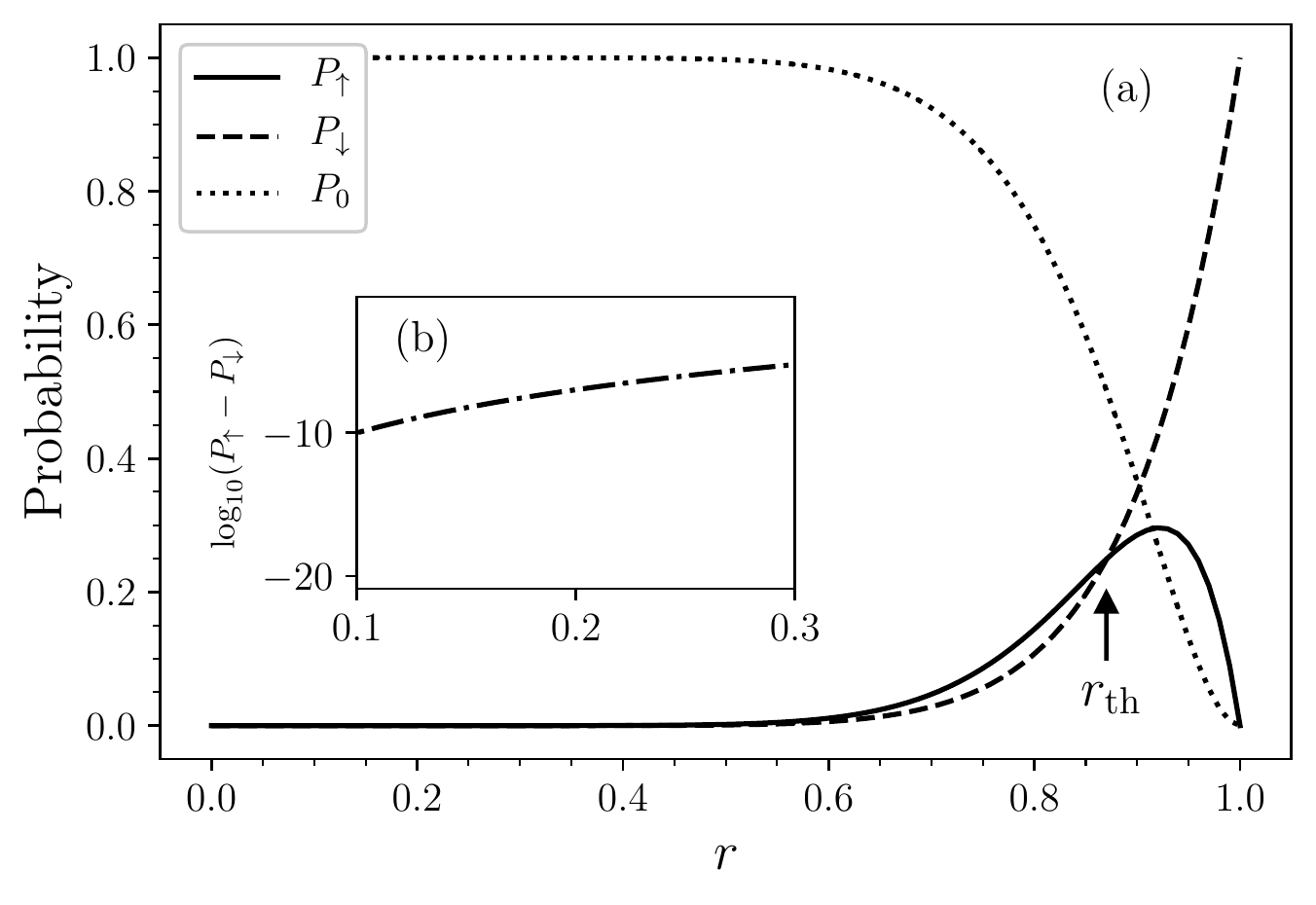}
\caption{\label{fig:DSMC_up_down} 
Main set: (a) Probability of the tracer moving up ($P_\uparrow(r; \beta)$), down
($P_\downarrow(r)$) or not moving ($\pz(r; \beta)$) for $N=10$
and $\beta=1.5 \bc$ (TA regime). 
The threshold radius is identified by $\pu=\pd$.
For $r>\rth$, we have $\pd>\pu$.
For $r<\rth$, we have $\pd<\pu$ (see the inset (b) for a closeup of the difference between the two).
The slowdown of the dynamics with small $r$ is encoded in $\pz$ going to 1.
}
\end{figure}

Note that the threshold is an attractor of the dynamics, in the sense that the tracer's distance from the center tends to shrink when $r>\rth$, and to expand when $r<\rth$ (Fig.~\ref{fig:DSMC_up_down}).
This can be seen clearly from Fig.~\ref{fig:DSMC_up_down}, where we show that $\pu>\pd~\forall r<\rth$, and $\pu<\pd~\forall r>\rth$. We also show $\pz(r;\beta)$, that goes to 1 as $r$ decreases and can be used as an indicator of the slowness of the dynamics.

In the entropic aging regime the tracer is attracted to the center of the sphere,
which it approaches over an infinitely long amount of time.
In the thermally activated regime the system fluctuates around $0<\rth<1$, but as time passes it becomes increasingly probable that low-energy configurations are reached, where the system will spend very long times before rising to the threshold again.
Finally, in the equilibrium phase the system is squeezed on the surface of the hypersphere [$r=1-O(1/N)$], and when low-energy configurations are reached the thermal agitation is strong enough to allow for the system to quickly go back  to the surface of the hypersphere.

\begin{figure}[tb]
    \centering
    \includegraphics[width=1.0\columnwidth]{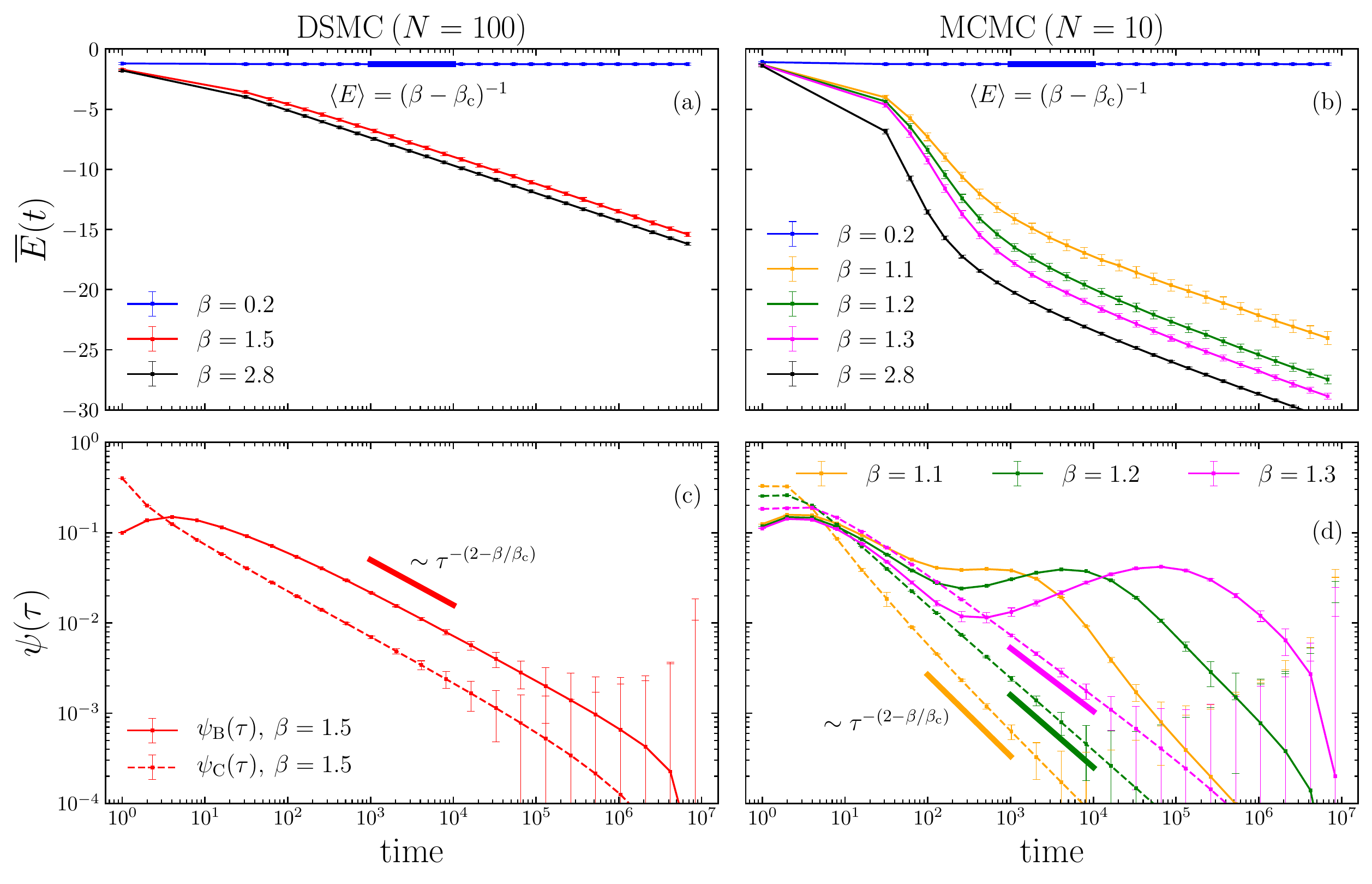}
    \caption{
    Top: We show the energy of three trajectories as a function of the Monte Carlo timestep $t$, for $N=100$ and $\delta=0.1$, each in a different dynamical regime. On the left we show DSMC dynamics (a), and on the right we show MCMC dynamics (b).
    The EP curve ($\beta=0.2$) converges quickly to its equilibrium value $\langle E(\beta)\rangle=-1.25$ (horizontal blue dashed line), while TA and EA curves ($\beta=1.5$ and 2.8, respectively) diverge logarithmically to $-\infty$.
    Bottom:
    Configuration and basin trapping time distributions, $\psi_\mathrm{C}(\tau_\mathrm{C})$ (dashed) and $\psi_\mathrm{B}(\tau_\mathrm{B})$ (solid), for DSMC (c) and MCMC (d) dynamics. Results are averaged over 200 identical simulations,
    each with 500 tracers.
    }
    \label{fig:psi}
\end{figure}

The described dynamical scenario is also encountered in the SM~\cite{barrat:95,bertin:03,bertin:12}. This is due to the combination of two ingredients: on one side the density of states $g(E)$ in Eq.~\eqref{eq:dos} is the same of the SM, and on the other the DSMC algorithm samples directly from $g(E)$. As a consequence, the dynamical succession of energies in our funnel is statistically the same as that of the SM, so we note that several results from the SM are realized in
DSMC dynamics. 
For example, in the SM, the distribution of persistence times in a configuration, $\psi_\mathrm{C}(\tau_\mathrm{C})$ (distribution of times spent in a configuration), and in a basin, $\psi_\mathrm{B}(\tau_\mathrm{B})$ (distribution of times spent under the threshold)
both decay as~\cite{bertin:12,cammarota:15}
\begin{equation}\label{eq:mu}
\psi(\tau)\sim1/\tau^{1+\mu}\,,
\end{equation}
with $\mu=2-\beta/\bc$,
which is what we find in our high-dimensional funnel (Fig.~\ref{fig:psi}, left bottom), and the energy decays logarithmically (Fig.~\ref{fig:psi}, left top).

Note, however, that finite-size effects are now different, because in network models such as TM, SM and REM, the lowest available energy depends on $N$. Therefore, for any finite $N$, $\overline{E}(t)$ eventually saturates, whereas in our model the energy decreases to $-\infty$ at any system size, so the dynamical phase diagram~\cite{gayrard:18b} can differ.

\subsection{Markov Chain Monte Carlo}
Especially given that we are introducing spatial effects in the dynamics of network models, it is perhaps more interesting to study also local dynamics.
We analyze a Markov Chain sampling of our $N$-dimensional funnel. 
From every point $\vecx_t$ in the hypersphere, a new point $\vecx_{t+1}$ is proposed by making a Gaussian shift
\begin{equation}\label{MCMC update scheme}
    \vecx_{t+1} = \vecx_{t} + \mathbf{\Delta},
\end{equation}
where $\mathbf{\Delta} \sim \mathcal{N}_N(0,\delta^2)$ is an $N$-dimensional Gaussian random variable with variance $\delta^2$ (meaning a diagonal covariance matrix with $\delta^2$ in each
entry) centered at the origin and randomly sampled at every time step.
The move is accepted with the Monte Carlo rate in Eq.~\eqref{eq:pmc}. The initial
configuration is uniformly drawn from the radius 1 hypersphere. 

Also in this case, we find the same three dynamical regimes that we found for the DSMC.
Since the equilibrium properties are independent of the type of dynamics (provided that it obeys detailed balance), we still have an EP for $\beta<\bc$. For higher $\beta$, there is a TA regime defined by the presence of a positive threshold radius. In Appendix~\ref{app:MCMC} we show that the TA regime terminates at $\beta=2$, where the dynamics is not intermittent anymore, and one reaches an EA regime, where the energy decreases steadily.

\begin{figure}[tbh]
    \centering
    \includegraphics[width=\columnwidth]{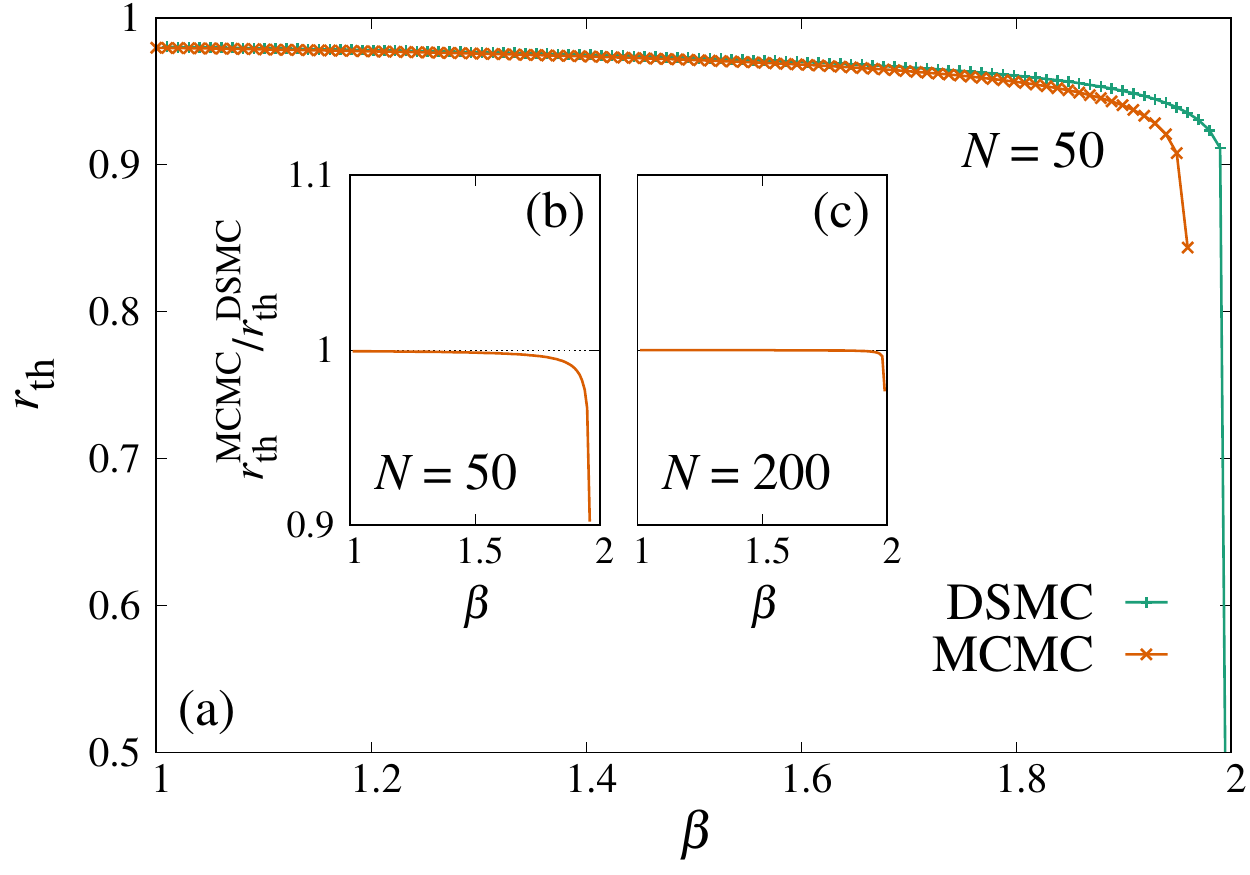}
    \caption{Main set: (a) Threshold radius as a function of $\beta$ in a system of size $N=50$, for DSMC [Eq.~\eqref{eq:rth}] and MCMC [Eq.~\eqref{eq:rth-beta}] dynamics. The $y$ axis is truncated at 0.5 to improve the figure's clarity. Points are shown in the curves in order to make overlapping curves visible.
    Insets: Ratio between the MCMC and the DSMC threshold radii for (b) $N=50$ and (c) $N=200$.}
    \label{fig:rth-dsmc-mcmc}
\end{figure}

We can derive the threshold radius that is valid for large $N$ by imposing that the probability of increasing and decreasing the energy is equal,
\begin{equation}\label{eq:cond-MCMC}
\pd(\rth;\delta)\equiv\pu(\rth;\beta,\delta)\,,
\end{equation}
where now Eq.~\eqref{eq:cond-MCMC} also accounts for the size of the MCMC step, $\delta$, in the upwards and downwards probabilities, which can be formally written as 
\begin{equation}\label{eq:pdownmc}
    \pd(x_0;\delta) \propto \int_{0<|\vecx|<x_0} d\vecx\, e^{-\frac{(\vecx-\vecx_0)^2}{2\delta^2}} \,, 
\end{equation}
and
\begin{equation}\label{eq:pupmc}
    \pu(x_0;\beta,\delta) \propto \int_{x_0<|\vecx| < 1} d\vecx\, e^{-\frac{(\vecx-\vecx_0)^2}{2\delta^2}}  e^{-\beta \Delta E(\vecx;\vecx_0)}\,.
\end{equation}
In Appendix~\ref{app:MCMC} we solve Eq.~\eqref{eq:cond-MCMC}, see that $\rth$ is independent from $\delta$ up to $o(1/N)$ terms, and derive an explicit form approximately identical to the DSMC one:
\begin{equation}\label{eq:rth-beta}
    r_\mathrm{th} (\beta) = \left(  \frac{2\bc -\beta}{\bc} \right)^{\frac{\bc}{N\left(\beta - \bc \right)}}
    + o\left(\frac1N\right)\,,
\end{equation}
In the large-$N$ limit the threshold radius for MCMC and DSMC coincides (see also Appendix~\ref{app:MCMC} and Fig.~\ref{fig:rth-dsmc-mcmc}). Further, both $\overline{E}(t)$ and $\psi(\tau)$ behave the same at long times (see Fig.~\ref{fig:psi}, right). 

\section{Trap-like behavior}\label{sec:trap}
In Ref.~\onlinecite{cammarota:15}, it was shown that the SM displays an activated aging dynamics that is effectively like that of the TM. In order to do so, we studied the time evolution of the energy, and defined energy basins dynamically, as the periods of time that the system remains at $E<\Eth$. The distributions of trapping times are shown in Fig.~\ref{fig:psi}--bottom.

We can use the exponent $\mu$ [Eq.~\eqref{eq:mu}] to show that the funnel model has TM dynamics, as was done in Ref.~\onlinecite{cammarota:15} for the SM, by studying the aging function $\Pi_\mathrm{B}(t_\mathrm{w},t_\mathrm{w}+t)$, defined as the probability of not changing basin between the times $t_\mathrm{w}$ and $t_\mathrm{w}+t$.\footnote{Details on definition and computation of $\psi(\tau)$ and $\Pi_\mathrm{B}(t_\mathrm{w},t_\mathrm{w}+t)$ are given in
Appendix~\ref{app:sim}.} 
To define the basins' threshold we used Eq.~\eqref{eq:rth} for both DSMC and MCMC.

\begin{figure}[tbh]
    \centering
    \includegraphics[width=1.0\columnwidth]{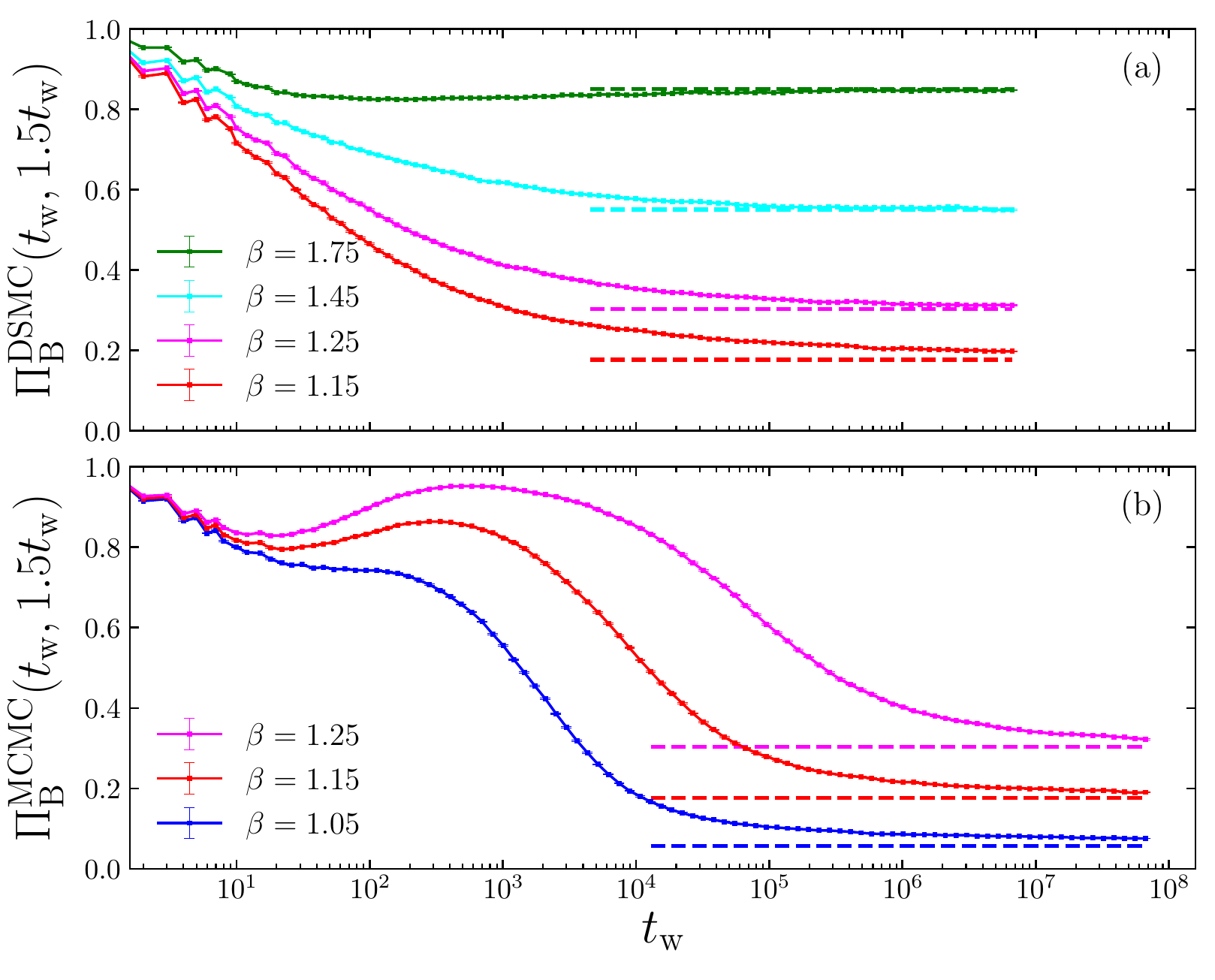}
    \caption{
    Top: (a) Aging functions $\Pi_\mathrm{B}(\tw,1.5\,\tw)$ for DSMC dynamics, with $N=100$, $\beta=1.15, 1.25, 1.45, 1.75$.
    Bottom: (b) Aging functions $\Pi_\mathrm{B}(\tw,1.5\,\tw)$ for MCMC dynamics for $N=10$ and $\beta=1.05, 1.15, 1.25$. The dashed horizontal lines
    correspond to the trap value $H_{2-\beta}(0.5).$ Results are averaged over 200 identical simulations, each with 500 tracers.}
    \label{fig:trap}
\end{figure}

In the TM, the aging function has a well-defined limiting value which depends only on the exponent $\mu$ [Eq.~\eqref{eq:mu}] and on the ratio $w=t/\tw$~\cite{bouchaud:92, bouchaud:95},
\begin{equation}\label{aging function}
    H_\mu(w) = \frac{\sin(\pi \mu)}{\pi}\int_w^\infty\frac{du}{(1+u)u^\mu}\,. 
\end{equation}
In Fig.~\ref{fig:trap}--top we show that the aging function in the TA regime converges clearly to the TM prediction in DSMC dynamics. The same is valid for MCMC dynamics which, being local, is much slower than DSMC, so our simulations are restricted to lower $N$ and $\beta$.

\begin{figure}[tbh]
    \centering
    \includegraphics[width=1.0\columnwidth]{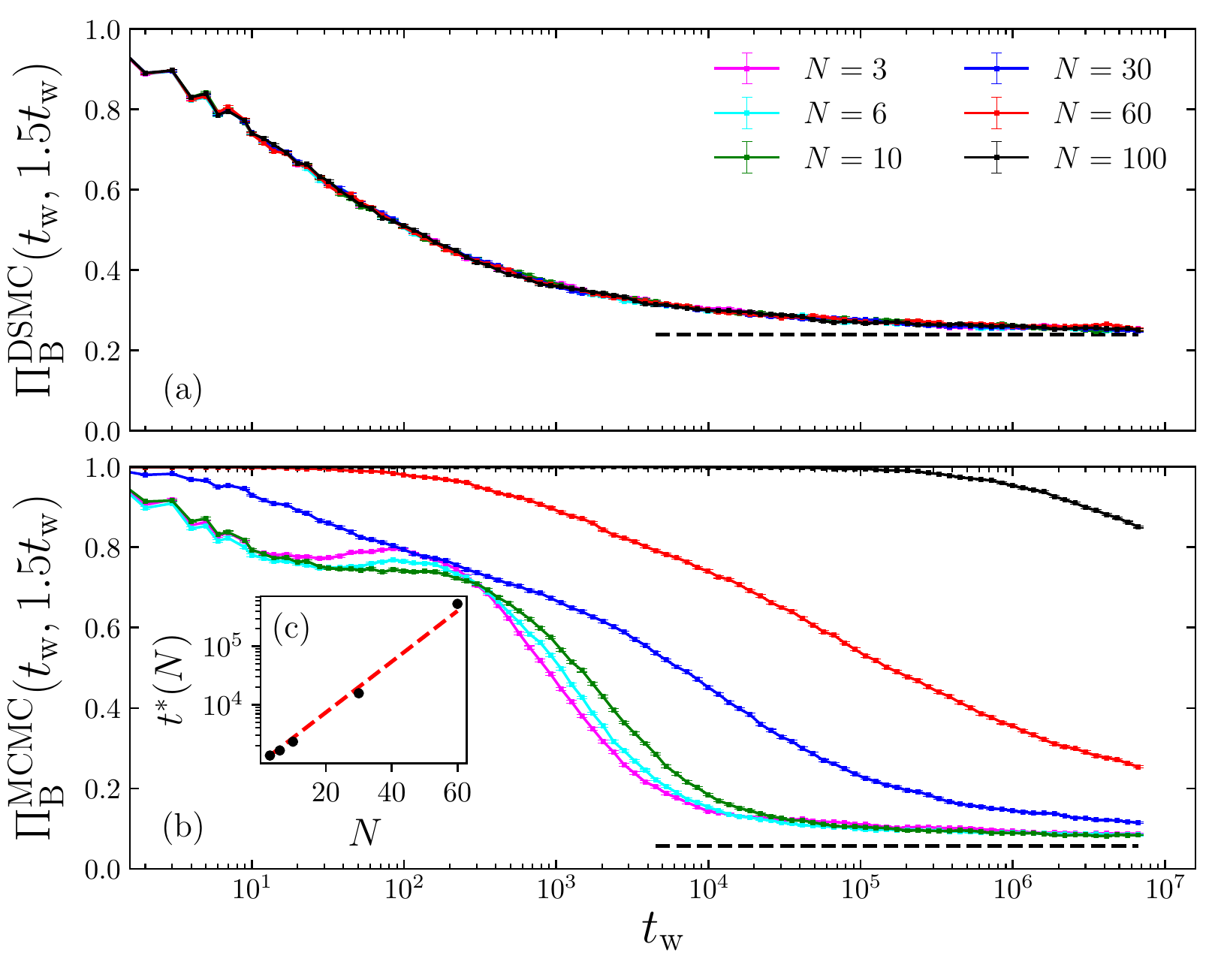}
    \caption{
    Top: (a) Aging functions $\Pi_\mathrm{B}(\tw,1.5\,\tw)$ for DSMC dynamics with $\beta=1.2$ and $N=3, 6, 10, 30, 60, 100.$
    Bottom: (b) Aging functions $\Pi_\mathrm{B}(\tw,1.5\,\tw)$ for MCMC dynamics with $\beta=1.05$ and the same values for $N.$
    The dashed horizontal lines correspond to the trap values $H_{2-\beta}(0.5).$ Results are averaged over
    100 identical simulations, each with 200 tracers. Inset: (c) The time at which the average $\Pi_\mathrm{B}^\mathrm{MCMC}(\tw,1.5\,\tw)$ function crosses 0.4 as a function of $N$ (obtained through linear interpolation). Note
    the semi-log scale on the $y$-axis implies exponential scaling in $N.$
    }
    \label{fig:finite-size}
\end{figure}

The strong slowing down in the MCMC dynamics can be appreciated from Fig.~\ref{fig:finite-size}.
As one can expect, the global update dynamics has no finite-size effects, whereas the local dynamics is increasingly slower as the system size increases.\footnote{This is true if $\delta$ is independent from the system size. If $\delta$ increases with the system size (which is not the case in typical local algorithms of many-body systems), the dependence on $N$ can be suppressed. 
To obtain that the ratio of the volume of configurations accessible in one step, divided by the total volume, should stay constant. We can use Eq.~\eqref{eq:pdown} to obtain $\delta\sim e^{-\frac{\mathrm{const.}}N}$.}
This slow down is exponentially large with the system
size (Fig.~\ref{fig:finite-size}, inset), which suggests that our funnel model is correctly capturing the nature of activated processes.

\section{Discussion \& Conclusions}\label{sec:conc}
We investigated the out-of-equilibrium dynamics of a tracer in an $N$-dimensional funnel landscape. The dynamics is dominated by the competition between an energetic pull towards the center, and an entropic push outwards due to the dimension of the space, which turns out to be equivalent to increasing the thermal noise by a factor $N$. As a consequence, the energetic contribution needs to scale with $N$ (or the temperature needs to be rescaled by $1/N$) for it to be relevant.

The properly rescaled model has a high-temperature equilibrium phase, and two low-temperature out-of-equilibrium regimes: a thermally activated regime in which the tracer intermittently comes up to the surface even though, on average, its energy decreases indefinitely, and an entropic aging regime in which this intermittency disappears. This same phenomenology is found in the Step Model (SM), a network model with random energies and no notion of distance~\cite{barrat:95,bertin:10,bertin:12}, which in our model is recovered in the limit of maximally delocalized and uncorrelated updates.

We find that, besides the system size dependence, this non-equilibrium behavior is independent of the chosen dynamics. We examined a global update method, Direct Sampling Monte Carlo (DSMC), and a local update, Markov Chain Monte Carlo (MCMC), which turned out to be equivalent, providing an example of the equivalence of equilibrium algorithms in out-of-equilibrium contexts. Extensions to other kinds of physical dynamics would arguably give the same results.

Unlike the SM, quenched disorder is not needed in order to have glassy TM-like activated dynamics. This is understood by comparing the SM with our funnel model with DSMC dynamics: the randomness due to disorder in the SM can be incorporated into that due to thermal fluctuations, giving the same kind of long-time activated dynamics.

The funnel model can be seen as an extension of the SM to a continuous landscape, where a notion of space, distance and dimension are now well-defined. This makes it viable to extend the TM paradigm (or the suitable modifications of it) to more realistic models, such as structural glasses.\footnote{
One may argue that most models of structural glasses enjoy translation invariance, which the funnel model lacks. However, even though we do not claim that this funnel model as is can faithfully represent a glass, translation invariance, or lack thereof, does not present with cause for concern. Indeed, translation invariance is not necessarily a feature of glasses, and breaking it often leads to an increased glassiness. For example, randomly pinning particles in a supercooled liquid breaks translation invariance, but exacerbates the glassy behavior~\cite{cammarota:12,karmakar:13}. Both in structural and in spin glasses, symmetries such as translation and rotation give rise to Debye modes in the density of states, which are not a signature of glassiness. To the contrary, the glassy low-frequency modes emerge when those symmetries are explicitly broken~\cite{angelani:18,baityjesi:15b}.}

A critique suffered by models such as TM and SM is that the excessive simplicity of their phase space makes it impossible to use them to describe any Hamiltonian system in realistic terms, which is solved by our funnel model. An alternative approach to associate the TM to models with microscopic degrees of freedom was proposed in Ref.~\onlinecite{junier:04}, by reformulating the number partitioning problem (NPP) as a Mattis spin glass, and focusing on single spin flip dynamics. This TM-like behavior is due to the presence of few low-energy configurations that are one spin flip away from typical energies. By changing the dynamics to multiple spin flips, the TM behavior disappears, and the dynamics mimics the SM when all the spins are updated simultaneously~\cite{junier:04b}. This is conceptually different from what we find for two main reasons:
\begin{itemize}
    \item In Ref.~\onlinecite{junier:04}, the TM dynamics is due to an equivalence at a level of the \emph{landscape}, which is composed by rare, point-like, regions with a very low energy in an environment where the system can otherwise move freely.
    Instead, in the funnel model (and in the SM), the origin of the TM-like behavior is purely due to entropy, and it is effective in the sense that the trapping time distributions are different from those of the TM, but their relationship with the aging functions is the same as that of the TM. We thus have two very different kinds of TM-like dynamics, one which is energy-driven, and another which is entropy-driven, and they should be treated differently. Defining basin hopping through the dynamics has also been done in experiments of glass-formers by looking at particle movement~\cite{fris:11}.
    However, the possible entropic origin of the observed activated dynamics is either dismissed, in favor of energy-based arguments, or it is incorporated into kinetic constraint arguments~\cite{vogel:04}.
    Both kinds of barriers (energetic and entropic) induce logarithmically slow dynamics~\cite{heuer:08}, and in realistic systems there is likely competition (or synergy) between the two kinds of effects, due to the presence of a collection of deep wide minima~\cite{baityjesi:18c}.

    \item We find a TM-like behavior for both local and global dynamics. The type of activation analyzed in the NPP is energy-driven~\cite{junier:04}, and in the limit of global updates the model resembles the SM~\cite{junier:04b}, which exhibits entropic TM-like activation~\cite{cammarota:15}. Our analysis suggests, therefore, that entropy-driven activation is more robust to changes in the dynamics, and that the NPP is likely to exhibit both energetic and entropic trap-like behaviors at the same time. We highlight these simple models and their activated behaviors in relation to their dynamics in Table~\ref{tab:params}.
\end{itemize}

\begin{table}[!htb]
\caption{\label{tab:params} For each model and dynamics, we specify whether we have energy- or entropy-driven activation.
Trap model (TM) and Step model (SM) live in a fully connected phase space, so they cannot have local dynamics. 
The funnel model studied in this paper consists of a single well, so it cannot have energy-driven trap-like behavior. It does have entropy-driven activation for both local and global dynamics.
The number partitioning problem (NPP) and the random energy model (REM) have energy-driven trap-like activation when using single spin flip dynamics. With global dynamics, the NPP has (as well as the exponential version of the REM)
entropy-driven activation.
This makes the NPP and REM models good candidates for having both kinds of trap-like behaviors simultaneously (i.e. with the same dynamics), since we can expect the same phenomenology of the funnel model.
}
\begin{ruledtabular}
\begin{tabular}{lcccc}
& \multicolumn{2}{c}{Global} & \multicolumn{2}{c}{Local}\\
\textrm{Model}&
\textrm{Energy}&\textrm{Entropy}& \textrm{Energy}&\textrm{Entropy}\\
\colrule
TM & \textbf{Yes} & No & No\footnote{\label{f1}
The trap and step models cannot have local dynamics.} & No\footnoteref{f1} \\
SM & No & \textbf{Yes} & No\footnoteref{f1} & No\footnoteref{f1} \\
Funnel & No & \textbf{Yes}\footnote{\label{f3}The subject of this work.} & No & \textbf{Yes}\footnoteref{f3} \\
NPP & No & \textbf{Yes} & \textbf{Yes} & ?\,\footnote{\label{f2} Models like NPP and REM could have entropically-activated
trap-like behavior also with local dynamics.}  \\
REM & No & Yes\footnote{An exponential REM with global dynamics is a SM. The usual, Gaussian, REM has not been examined.} & \textbf{Yes} & ?\,\footnoteref{f2} \end{tabular}
\end{ruledtabular}
\end{table}

Our funnel model introduces Euclidean space in the SM, and shows an alternative way of introducing locality, providing the possibility of local moves but without the multiplicity of local minima that characterize energy-driven trap landscapes.
As it also happens for the NPP~\cite{junier:04b}, the SM is only the limit for maximally global dynamics of our model, but now locality is different than in the NPP, since it involves a Euclidean metric, and therefore, finite-size effects are also different.
In fact, in the SM and the other aforementioned lattice models,
$N$ determines the lowest reachable energy, whereas here $N$ is related to the amplitude of the noise. This implies that SM and funnel model are the same model only in the $N\to\infty$ limit using DSMC dynamics. Another way to see this is that, even for finite $N$, in the funnel model, the ground state is at $-\infty$, and it is almost impossible for any algorithm with any amount of noise to reach the center of the hypersphere. A direct consequence of this is that, unlike TM, SM, REM and NPP, a finite-size funnel model at $\beta>\bc$ will never reach equilibrium.
Furthermore, the introduction of local dynamics and the connection to particle systems uncover that entropic TM-like aging also displays an exponential slowing down with the system size, which is a fundamental trait of activation which needed to be observed.

The interaction potential to which these particles are subject, although exotic, can be found in several situations, such as in bosonic systems~\cite{weiss:04}, 
when converting into extensive problems with an exponential scaling in the system size~\cite{junier:04}, or by reformulating number theory problems in terms of a cost function~\cite{junier:04}.
An interesting development of our work would be the exploration of entropy-driven activation in $r^\alpha$ potentials (for example the case $\alpha=-1$ would represent $N$ particles in a Coulomb potential). Such further directions will be the subject of future work.

\begin{acknowledgments}
MBJ and MRC are very grateful to D. R. Reichman for supporting them in doing independent research in his lab. MBJ thanks S. Franchini for pointing out the connection to bosonic systems.
This work was funded by the Simons Foundation for the collaboration “Cracking the Glass Problem” (No. 454951 to D. R. Reichman). 
MRC acknowledges support from the US Department of Energy through the Computational Sciences Graduate Fellowship (DOE CSGF) under Grant No. DE-FG02-97ER25308.
MBJ thanks MINECO for partial support through research contract No  PGC2018-094684-B-C21 (contract partially funded by FEDER).
\end{acknowledgments}

\appendix
\section{Simulation Details}\label{app:sim}
In this Appendix we explain our simulation procedures, and explain our parallel code, that we provide for open access at \code{https://github.com/x94carbone/hdwell}.

For each choice of the parameters and dynamics we simulate $M$ batches (usually 100 to 200) of trajectories of $\tau_\mathrm{max}=10^7$ to $10^8$ time steps.
Each of the $m$ tracers per batch (usually 200 to 500) is computed in parallel, and are used to compute distribution averages effectively. From each batch we obtain the whole curves $\psi(\tau)$ and $\Pi(\tw,\tw+t)$ which can then be averaged among batches. The specific details of how these values are calculated are summarized in this Appendix.

\subsection{Monte-Carlo Procedure}
The details of the simulation algorithms are henceforth summarized:
\begin{enumerate}
    \item Initialize on the $N$-dimensional sphere with a uniform distribution (so for large $N$ the tracer is initially at $r\simeq1$). To sample uniformly the hypersphere we use the algorithm in Ref.~\onlinecite{barthe:05}.
    \item For each tracer at time $t$, make a proposal move towards $\vecx_{t + 1}^*$  as follows:
    \begin{enumerate}
        \item If DSMC: $\vecx_{t + 1}^*$ uniform in the hypersphere (using Ref.~\onlinecite{barthe:05}).
        \item If MCMC: $\vecx_{t + 1}^*$ according to Eq.~\eqref{MCMC update scheme}.
    \end{enumerate}
    \item Accept or reject the move with the Metropolis rule Eq.~\eqref{eq:pmc}.
    \item If the timestep is designated for recording quantities of interest, save the value of the energy, $\psi$ and $\Pi$-values for each tracer.
    \item Update the timestep: $t \leftarrow t + 1.$
    \item Repeat 2-5 until $t = t_\mathrm{max}.$ \\
\end{enumerate}

\subsection{Calculation of the Trapping Time Distributions}
To compute $\psi_\mathrm{C},$ the following procedure is used: A counter is initialized for each tracer in a simulation which keeps track of the number of time steps that tracer remains in a single configuration.
Since this is a continuous landscape, the distance from the center of the well is a sufficient proxy for
the exact configuration, since we can neglect the probability of changing configuration maintaining exactly the same radius.
Therefore, the trapping time is measured as the number of steps during which the system is at the same $r$.
At every time step, the configuration of each tracer is
queried. If $r_{t} = r_{t + 1},$ that tracer's counter increases by 1.
If $r_{t} \neq r_{t + 1}$, we immediately update a histogram (with $\log_2$-spaced bins) with the the value of the trapping time. This procedure allows for a sizable reduction of the memory devoted to the measurements~\cite{baityjesi:18}.

A similar procedure is used to calculate the values for $\psi_\mathrm{B}.$ A separate counter keeps track of the number of time steps that a tracer is below $\Eth$ in a basin. As soon as the tracer rises above $\Eth$ this counter is logged and reset in the same way we described for $\psi_\mathrm{C}$.\\

\subsection{Calculation of the Aging Functions}
Finally, we make note of how we calculate values for $\Pi_\mathrm{B}$ during the simulation. Note that
the calculation of $\Pi_\mathrm{C}$ is analogous, where instead of the basin index described further on, the
configuration proxy $r_t$ is used, in the same way that we described for $\psi_\mathrm{C}$.
The quantity $\Pi_\mathrm{B}$ is the probability of \emph{not} changing basin between two times. Stated another way, $\Pi_\mathrm{B}(\tw, \tw+t)$ is the probability, being the tracer in some basin at $\tw,$
that it is in the \emph{same} basin at $(t + \tw)=\tw(1+w)$ 
 that the tracer is in that same basin (having not left).
In this work we take $w=0.5$.

To keep track of the particular basin a tracer is in, a basin index $\mathcal{B}_j$ is
kept for every tracer $j$ and has the following properties.
\begin{enumerate}
    \item If at time step $\tw,$ tracer $j$ is in its $n^\mathrm{th}$ basin (meaning it has entered
    and left $n-1$ basins before $\tw$), then $\mathcal{B}_j = n.$
    \item If the tracer has just left its $n^\mathrm{th}$ basin at $\tw,$ then
    $\mathcal{B}_j = n + i,$ where $i$ is the imaginary number. The choice of using
    complex numbers to index whether a tracer is in or out of a basin is arbitrary, but
    allows for simpler notation in the code.
    \item When the tracer reenters a basin, the imaginary component of $\mathcal{B}_j$ is
    set back to 0, and the real part increments: $n \gets n + 1.$
\end{enumerate}
Thus in summary, the real part of $\mathcal{B}_j$ references the index of the last basin
that tracer was in, and the presence of an imaginary component is used to index
whether or not that tracer is currently in or out of a basin. 
If $\Im\{\mathcal{B}(\tw)\}\neq0$, the measurement is discarded in computing the
normalization of $\Pi_B$, since the tracer is initially not in a basin. 
If instead $\Im\{\mathcal{B}_j(\tw)\}=0$, then for a particular tracer, 
\begin{itemize}
    \item if $\mathcal{B}_j(\tw)=\mathcal{B}_j(\tw+t)~~\Rightarrow~~\Pi_B(t,t')=1$ 
    \item if $\mathcal{B}_j(\tw)\neq\mathcal{B}_j(\tw+t)~~\Rightarrow~~\Pi_B(t,t')=0$ 
\end{itemize}{}
As we described for the trapping time distributions, we extract a curve $\Pi(\tw,\tw(1+w))$ from each batch of runs, and compute statistical error bars by comparing batches.

\begin{widetext}

\section{MCMC calculation of the threshold}\label{app:MCMC}
To evaluate the threshold radius in the MCMC approach we need to solve the equality 
\footnote{In this appendix we set $\bc=1$.}
\begin{equation}
\pd(\rth;\delta)=\pu(\rth;\beta,\delta)\,,
\end{equation}
with
\begin{equation}\label{eq:pdownmc2}
    \pd(x_0;\delta) =   N_{\delta} \int_{0<|\vecx|<x_0} d\vecx\, e^{-\frac{(\vecx-\vecx_0)^2}{2\delta^2}} \, , \qquad \pu(x_0;\beta,\delta) = N_{\delta} \int_{x_0<|\vecx| < 1} d\vecx\, e^{-\frac{(\vecx-\vecx_0)^2}{2\delta^2}}  e^{-\beta \Delta E}\,,
\end{equation}
where we defined the normalization constant
\begin{equation}\label{eq:normmc}
    N_{\delta} = \left(\int_{0\leq|\vecx|<1} d\vecx\, e^{-\frac{(\vecx-\vecn)^2}{2\delta^2}} \right)^{-1} \, , \quad |\vecn|=1\,.
\end{equation}
We can express the integrals in spherical coordinates as
\begin{equation}\label{commint}
\int d\vecx\, e^{-\frac{(\vecx-\vecn)^2}{2\alpha^2}}
 = \int d|\vecx|\, |\vecx|^{N-1}\, d\Omega^{N} e^{-\frac{\left( |\vecx|^2 + 1 - 2|\vecx|\cos\theta \right)}{2\alpha^2}} \,,
\end{equation}
where $\theta$ is the angle between $\vecx$ and $\vecn$ and $\Omega^N$ is the solid angle in $N$ dimensions. The $\theta$-integration can be singled out,
\begin{equation}
    \int d\Omega^{N} e^{\frac{|\vecx|\cos\theta}{\alpha^2}} = \tilde{\Omega}^{N-1} \int_{-1}^{1}  e^{\frac{|\vecx|\cos\theta}{\alpha^2}} d\cos\theta  =  \frac{\tilde{\Omega}^{N-1} \alpha^2}{|\vecx|} \left( e^{\frac{|\vecx|}{\alpha^2}} - e^{-\frac{|\vecx|}{\alpha^2}}  \right)\,,
\end{equation}
and the last result can be substituted back into Eq.~\eqref{commint} to obtain
\begin{equation}\label{eq:red}
    \int d\vecx\, e^{-\frac{(\vecx-\vecn)^2}{2\alpha^2}} =  \tilde{\Omega}^{N-1} \alpha^2 \int d|\vecx|\, |\vecx|^{N-2}\,  \left(e^{-\frac{(|\vecx|-1)^2}{2\alpha^2}} - e^{-\frac{(|\vecx|+1)^2}{2\alpha^2}} \right)\,,
\end{equation}
where $\tilde{\Omega}^{N-1}$ is the result of the integration of the residual angular coordinates, and it is independent of the radial position. 
Rescaling the integrand in Eq.~\eqref{eq:pdownmc2} we obtain
\begin{equation}
    \pd(x_0;\delta) =   x_0^N \, \left(\frac{   \int_{0<|\vecx|<1} d\vecx\, e^{-\frac{(\vecx-\vecn)^2}{2\alpha^2}}  }{  \int_{0<|\vecx|<1} d\vecx\, e^{-\frac{(\vecx-\vecn)^2}{2\delta^2}}   } \right)  \, ;\qquad \alpha = \frac{\delta}{x_0}\,.
\end{equation}
Denoting $\pd(x_0)$ the corresponding probability in the DSMC, we have
\begin{equation}
    \pd(x_0;\delta) \geq \pd(x_0) \, , \qquad \lim_{x_0 \to 1} \pd(x_0;\delta) = \pd(x_0) \, .
\end{equation}
In addition we can use Eq.~\eqref{eq:red} to show that for $N\gg 1$ and $\delta \propto N^{-\xi}$, $0<\xi<1$ we have\footnote{We assume this scaling for $\delta$ and $N$ to be valid through the remaining of the appendix.}
\begin{equation}
    \lim_{x_0 \to 0} \pd(x_0;\delta) = \frac{2 }{\delta^2 }\,  \pd(x_0) \, \left[ 1 + O\left(N^{-1} \right) \right]\,.
\end{equation}
With similar manipulations we obtain an equivalent expression for $\pu$,
\begin{equation}
    \pu(x_0;\beta, \delta) = x_0^{N} \,  \left(\frac{   \int_{1<|\vecx|<\frac{1}{x_0}} d\vecx\, |\vecx|^{-\beta N} \, e^{-\frac{(\vecx-\vecn)^2}{2\alpha^2}}  }{  \int_{0<|\vecx|<1} d\vecx\, e^{-\frac{(\vecx-\vecn)^2}{2\delta^2}}   } \right) \,.
\end{equation}
or, integrating out the angular coordinate,
\begin{equation}
    \pu(x_0;\beta, \delta) = x_0^{N-2} \,  \left(   \frac{   \int_{1}^{1/x_0} dy \, y^{N-2-\beta N} \, \left(  e^{-\frac{(y-1)^2}{2\alpha^2}} - e^{-\frac{(y+1)^2}{2\alpha^2}} \right) }{  \int_{0}^{1} dy \, y^{N-2} \left(  e^{-\frac{(y-1)^2}{2\delta^2}} - e^{-\frac{(y+1)^2}{2\delta^2}} \right)   }  \right)\,.
\end{equation}
We can relate $\pu(x_0;\beta,\delta)$ with $\pu(x_0;\beta)$, the probability of increasing the radius in the DSMC. In fact, it is easy to show that
\begin{equation}
    \lim_{x_0 \to 0} \pu(x_0;\beta, \delta) = \frac{2 }{\delta^2 }\,  \pu(x_0;\beta) \, \left[ 1 + O\left(N^{-1} \right) \right]\,.
\end{equation}

\subsection{Threshold Radius for MCMC steps}
We are interested in the ratio between the two probabilities, which can be expressed as
\begin{equation}
    \frac{\pu(x_0;\beta, \delta)}{\pd(x_0;\delta)} = R(x_0;\beta, \alpha) =  \frac{   \int_{1<|\vecx|<\frac{1}{x_0}} d\vecx\, |\vecx|^{-\beta N} \, e^{-\frac{(\vecx-\vecn)^2}{2\alpha^2}}  }{  \int_{0<|\vecx|<1} d\vecx\, e^{-\frac{(\vecx-\vecn)^2}{2\alpha^2}}   } \,,
\end{equation}
or, keeping only the radial coordinates,
\begin{equation}
    R(x_0;\beta, \alpha) =    \frac{   \int_{1}^{1/x_0} dy \, y^{N-2-\beta N} \, \left(  e^{-\frac{(y-1)^2}{2\alpha^2}} - e^{-\frac{(y+1)^2}{2\alpha^2}} \right) }{  \int_{0}^{1} dy \, y^{N-2} \left(  e^{-\frac{(y-1)^2}{2\alpha^2}} - e^{-\frac{(y+1)^2}{2\alpha^2}} \right)   } \,. 
\end{equation}
In order to find the threshold radius, we need to solve
\begin{equation}\label{eq:rthR}
    R(\rth;\beta, \alpha) = 1\,.
\end{equation}{}
In the limit $x_0 \to 0$ we find the same value for the ratio as in the DSMC (in contrast with the single probabilities, the ratio has no $N^{-1}$ corrections):  
\begin{equation}
    \lim_{x_0 \to 0} R(x_0;\beta, \alpha) = \frac{1}{\beta -1}\,,
\end{equation}
which gives a threshold radius $r_{th}=0$ for $\beta = 2$. 
By imposing a null variation of $R(x_0;\beta, \alpha)$ with respect to $x_0$ and $\beta$ it is easy to show that the threshold radius is a decreasing function of $\beta$. Since $\rth = 0$ at $\beta=2$, there cannot be any entropically activated dynamics for $\beta>2$.
Using a saddle-point-like approximation the ratio $R(x_0;\beta, \alpha)$ reduces to
\begin{equation}\label{eq:R1}
    R(x_0;\beta, \alpha)  = \frac{\mathcal{N}(x_0;\beta, \alpha)}{D(\alpha)}\,,
\end{equation}
\begin{eqnarray}
    \mathcal{N}(x_0;\beta, \alpha)   \approx  \frac{\left(1-e^{-\frac{2}{\alpha^{2}}}\right)}{1+N \Delta \beta}\left[1-x_{0}^{1+N \Delta \beta }\right] +   \frac{2}{\alpha^{2}}e^{-\frac{2}{\alpha^{2}}}\frac{1-x_{0}^{N \Delta \beta}\left[1+\left(1-x_{0}\right)N \Delta \beta\right]}{N \Delta \beta\left[1+N \Delta \beta\right]} + \nonumber \\
    + \frac{1}{2\alpha^{2}}\left(\left(1-\frac{4}{\alpha^{2}}\right)e^{-\frac{2}{\alpha^{2}}}-1\right)\frac{2-x_{0}^{N \Delta \beta-1}\left(1-x_{0}\right)^{2}N^{2}\Delta\beta^{2}}{N \Delta \beta\left[N \Delta \beta+1\right]\left[N \Delta \beta-1\right]} \,,
\end{eqnarray}
\begin{equation}
    D(\alpha) \approx  \frac{\left(1-e^{-\frac{2}{\alpha^{2}}}\right)}{N-1}-\frac{2}{\alpha^{2}}e^{-\frac{2}{\alpha^{2}}}\frac{1}{N\left(N-1\right)}+\frac{1}{2\alpha^{2}}\left(\left(1-\frac{4}{\alpha^{2}}\right)e^{-\frac{2}{\alpha^{2}}}-1\right)\frac{2}{N\left(N+1\right)\left(N-1\right)}
\end{equation}
where we defined $\Delta \beta = \beta - 1 $. 
As long as $x_0 \gg \delta$ the last expressions simplify to
\begin{equation}\label{eq:Rx0ggdelta}
    R(x_0;\beta, \alpha) \approx \frac{\left(1-x_{0}^{1+N \Delta \beta }\right)}{\Delta\beta}  \left[  1 - \frac{1+\Delta\beta}{N \Delta\beta} - \frac{x_0^2}{N^2 \delta^2}\left(\frac{2-x_{0}^{N \Delta \beta}\left(1-x_{0}\right)^{2}N^{2}\Delta\beta^{2}}{2 \Delta \beta^2 \left(1-x_{0}^{N \Delta \beta }\right)} - 1 \right) \right] \,.
\end{equation}
Note that the condition $x_0 \gg \delta$, given the scaling of $\delta$ with $N$ that we assumed, is true in the whole sphere barring a negligible volume around the origin which becomes important only when $\beta \to 2$. In this regime however the MCMC dynamics is well approximated by the DSMC one.

In Eq.~\eqref{eq:Rx0ggdelta}, the last term in square brackets is sub-leading in $N^{-1}$, so the equation $R(\rth;\beta, \alpha) = 1$ can be solved perturbatively. At the lowest order in $N^{-1}$ we obtain a solution similar to the DSMC one,
\begin{equation}
    \rth^{(0)}\left( \beta \right) = \left(1- \Delta\beta \right)^{\frac{1}{N\Delta\beta +1}}\,.
\end{equation}
Plugging this term in $R(x_0;\beta, \alpha)$ we obtain
\begin{equation}
    R(x_0;\beta, \alpha) \approx \frac{\left(1-x_{0}^{1+N \Delta \beta }\right)}{\Delta\beta}  \left[  1 - \frac{1+\Delta\beta}{N \Delta\beta} - \frac{\left(1- \Delta\beta \right)^{\frac{2}{N\Delta\beta }}}{N^2 \delta^2}\left(\frac{2-\left(1- \Delta\beta \right)\left(1-\left(1- \Delta\beta \right)^{\frac{1}{N\Delta\beta}}\right)^{2}N^{2}\Delta\beta^{2}}{2 \Delta \beta^3} - 1 \right) \right]\,, 
\end{equation}
and
\begin{equation}\label{eq:rth-last}
    \rth\left( \beta \right) \approx \left( 1  -\Delta\beta \right)^{\frac{1}{N\Delta\beta + 1}} \left[ 1 -  \frac{1+\Delta\beta}{N^2 \Delta\beta \left( 1- \Delta\beta \right)} - \frac{\left(1- \Delta\beta \right)^{\frac{2}{N\Delta\beta}-1}}{N^3 \delta^2}\left(\frac{2-\left(1- \Delta\beta \right)\left(1-\left(1- \Delta\beta \right)^{\frac{1}{N\Delta\beta}}\right)^{2}N^{2}\Delta\beta^{2}}{2 \Delta \beta^3} - 1 \right)  \right]\,.
\end{equation}
To leading order Eq.~\eqref{eq:rth-last} coincides with the DSMC threshold radius, Eq.~\eqref{eq:rth}.

\end{widetext}

\bibliographystyle{apsrev4-1}

\end{document}